\documentclass[preprint,aps,amsmath,amssymb,nofootinbib,12pt]{revtex4}

\usepackage{epsfig}
\usepackage{slashed}
\usepackage{graphicx}
\usepackage{multirow,color}
\usepackage{amsmath}
\usepackage{float}
\usepackage{diagbox}
\usepackage{CJK}
\usepackage{color}
\usepackage{xcolor}
\usepackage{times}
\usepackage{subfigure}
\usepackage{bm}
\usepackage{braket}
\usepackage{booktabs}
\usepackage{array}
\usepackage[mathscr]{euscript}
\usepackage{caption}
\usepackage{makecell}
\usepackage{epstopdf}
\usepackage{hyperref}

\makeatletter

\newcommand{\Rmnum}[1]{\expandafter\@slowromancap\romannumeral #1@}
\makeatother
\graphicspath{{fig/}}
\begin{document}
\title{Probing the ALP--Photon--Dark Photon Interaction through the Mono-Photon plus Missing-Energy Signature at a Muon Collider}
\author{Chong-Xing Yue$^{1,2}$}
\thanks{cxyue@lnnu.edu.cn}
\author{Yang-Yang Bu$^{1,2}$}
\thanks{byy20011020@163.com~(Corresponding author)}
\author{Xin-Yang Li$^{1,2}$}
\thanks{lxy91108@163.com}
\author{Feng-Xia Tang$^{1,2}$}
\thanks{tfx280616@163.com}

\affiliation{
$^1$Department of Physics, Liaoning Normal University, Dalian 116029, China\\
$^2$Center for Theoretical and Experimental High Energy Physics, Liaoning Normal University,  Dalian 116029, China}

\begin{abstract}
The dark axion portal provides a well-motivated framework for new physics beyond the Standard Model~(SM), in which an axion-like particle~(ALP) couples simultaneously to a photon and a dark photon, leading to distinctive collider signatures. In this work, we investigate the sensitivity to the $a$--$\gamma$--$\gamma'$ interaction via the mono-photon channel arising from $\mu^{+}\mu^{-} \to a \gamma'$ with $a \to \gamma \gamma'$ at the $1.5~\mathrm{TeV}$ muon collider with an integrated luminosity of $\mathcal{L}=500~\mathrm{fb}^{-1}$ for the polarized and unpolarized beam configurations, $P(\mu^+,\mu^-)=(-100\%,+100\%)$ and  $P(\mu^+,\mu^-)=(0,0)$, using detector-level simulation. The signal is characterized by a single energetic photon accompanied by missing energy as the dark photons escape detection. The projected sensitivity reaches the $\mathcal{O}(10^{-5})~\mathrm{GeV}^{-1}$ level for ALP masses in the range $1~\mathrm{GeV}\leq m_a\leq1200~\mathrm{GeV}$ for $P(\mu^+,\mu^-)=(-100\%,+100\%)$, while for $P(\mu^+,\mu^-)=(0,0)$ the projected sensitivity reaches the $\mathcal{O}(10^{-4})~\mathrm{GeV}^{-1}$ level for the same ALP mass range. These results demonstrate that the mono-photon signature at a muon collider provides a sensitive probe of the $a$--$\gamma$--$\gamma'$ interaction over a wide mass range.
\end{abstract}
\maketitle

\section{Introduction}
The Standard Model~(SM) provides a remarkably successful description of particle physics, with its predictions confirmed by a wide range of experimental measurements. Nevertheless, it remains incomplete and leaves several fundamental questions unanswered. In particular, from a theoretical perspective, the SM does not provide satisfactory explanations for the gauge hierarchy problem~\cite{Feng:2013pwa}, while from a phenomenological standpoint it lacks a viable candidate for dark matter~(DM)~\cite{Duffy:2009ig,Chadha-Day:2021szb} and fails to account for the matter-antimatter asymmetry of the Universe~\cite{DiBari:2021fhs}, the origin of neutrino masses~\cite{Gonzalez-Garcia:2007dlo}, and other outstanding issues. In addition, the absence of gravity within the SM framework further highlights its limitations. These shortcomings strongly motivate the exploration of new physics beyond the SM. Among the various proposed extensions, weakly coupled light particles have attracted considerable attentions, as they can evade existing experimental bounds while still leading to observable signatures. Typical examples include axion-like particles~(ALPs) and dark photons, which naturally arise in many theoretical frameworks and provide well-motivated candidates for new physics.

ALP, denoted as $a$, is a CP-odd pseudo Nambu--Goldstone bosons associated with spontaneously broken global symmetries, and can be regarded as a generalization of the QCD axion originally proposed to solve the strong CP problem~\cite{Peccei:1977hh,Peccei:1977ur,Weinberg:1977ma,Dine:1981rt} via the Peccei--Quinn~(PQ) mechanism. Unlike the QCD axion, however, ALPs are not constrained by a fixed relation between their masses and couplings, leading to a significantly enlarged and more flexible parameter space. In addition, ALPs provide a framework for addressing phenomenological puzzles, for example as candidates for the cosmological dark matter (DM) relic density~\cite{Boehm:2003hm,Dolan:2014ska,Hochberg:2018rjs}, as mediators between visible and hidden sectors, and as probes of high-scale physics. Their couplings to gauge bosons, in particular, give rise to characteristic experimental signatures that can be tested in both low-energy and high-energy collider experiments.

Dark photon, commonly denoted as $\gamma'$, arises from an additional $U(1)_D$ gauge symmetry, which is a common feature in many extensions of the SM, such as grand unified theories and string-theoretic models~\cite{Fabbrichesi:2020wbt}. Its interactions with the SM particles can occur through kinetic mixing, leading to effective couplings between the dark photon and the electromagnetic current. Depending on its mass and coupling strength, the dark photon can give rise to a variety of observable signatures in collider experiments and other precision measurements. Its simple gauge structure and well-defined interaction mechanism make it a compelling target for experimental searches. Moreover, when combined with ALPs, the dark photon can participate in additional effective interactions that connect the visible and hidden sectors, leading to novel phenomenological implications.

While ALP and dark photon have been extensively studied individually, their interplay provides an additional and well-motivated avenue for probing new physics beyond the SM. In particular, the dark axion portal~\cite{Kaneta:2016wvf} introduces effective couplings among the ALP, the photon, and the dark photon. Compared to conventional ALP interactions with gluons, fermions, or electroweak gauge bosons~\cite{Phan:2023dqw,Anuar:2024qsz,Ghebretinsaea:2022djg,Mimasu:2014nea,Cheung:2024qge,Haghighat:2020nuh,Bauer:2018uxu,Zhang:2021sio,Wang:2022ock,Yue:2022ash,Yue:2021iiu,Cheung:2023nzg,Jiang:2024cqj,Gao:2024rgl}, the $a$--$\gamma$--$\gamma'$ interaction leads to qualitatively distinct phenomenology, not only in astrophysical and cosmological contexts~\cite{Kaneta:2017wfh,Choi:2018mvk,Kalashev:2018bra,Choi:2019jwx,Hook:2019hdk,Arias:2020tzl,Hook:2021ous,Domcke:2021yuz,Gutierrez:2021gol,Carenza:2023qxh,Hook:2023smg,Hong:2023fcy,Broadberry:2024pkv,DiazSaez:2024dzx,Arias:2025nub} but also at collider and reactor experiments that differ from those associated with conventional kinetic mixing scenarios, where it gives rise to characteristic signals~\cite{deNiverville:2018hrc,Biswas:2019lcp,Deniverville:2020rbv,Lane:2023eno,Jodlowski:2023yne,Jodlowski:2024ayf,Chen:2025hbh,Arias:2025tvd,Gninenko:2026mgn,Wang:2026fnr}. In addition, this interaction has been studied in various other contexts~\cite{Chen:2024jbr,Ding:2025eqq}.

Given that this interaction has been explored at electron--positron and hadron colliders but has not yet been systematically studied at muon colliders, it is well motivated to investigate the sensitivity of a muon collider and to assess its relative capability in probing the $a$--$\gamma$--$\gamma'$ coupling. Concretely, we investigate the process $\mu^+\mu^- \to a \gamma'$, followed by the decay $a \to \gamma \gamma'$, where the production is mediated by both the photon and the $Z$ boson, leading to a mono-photon signature accompanied by missing energy as the dark photons escape detection. This study is carried out at the muon collider operating at a center-of-mass energy of $\sqrt{s} = 1.5~\mathrm{TeV}$ with an integrated luminosity of $\mathcal{L} = 500~\mathrm{fb}^{-1}$ for the beam polarizations $P(\mu^+, \mu^-) = (-100\%, +100\%)$ and the unpolarized beam configuration $P(\mu^+, \mu^-) = (0,0)$, allowing us to assess the sensitivity to the $a$--$\gamma$--$\gamma'$ coupling and derive the corresponding projected sensitivity to the free parameters.

The paper is organized as follows. In Sec.~II, we introduce the theoretical framework of the $a$--$\gamma$--$\gamma'$ interaction together with the existing constraints. Under the assumption that the ALP mass is larger than that of the dark photon, $m_a > m_{\gamma'}$, we further present the partial decay widths of the ALP and discuss the corresponding branching ratios. In Sec.~III, we present a detailed analysis of the signal process $\mu^+\mu^- \to a \gamma'$ with $a \to \gamma \gamma'$ based on detector simulation at the $1.5~\mathrm{TeV}$ muon collider, and compare our results with existing experimental constraints and research findings from other colliders. Finally, we summarize our main results and conclusions in Sec.~IV.

\section{The theory framework}
The gauge-invariant Lagrangian for a light ALP with shift symmetry before electroweak symmetry breaking can be written as~\cite{Bauer:2018uxu,Bauer:2017ris}
\begin{eqnarray}
\begin{split}
\label{eq:2.1}
\mathcal{L} \supset
g_s^2 C_{GG}\, a\, G_{\mu\nu}^A \tilde{G}_A^{\mu\nu}
+g^2 C_{WW} \, a \, W_{\mu\nu}^A \tilde{W}_A^{\,\mu\nu} \\
+ g'^2 C_{BB} \, a \, B_{\mu\nu} \tilde{B}^{\mu\nu}
+ g_D'^2 C_{BB_D} \, a \, B_{\mu\nu} \tilde{B}_D^{\mu\nu},
\end{split}
\end{eqnarray}
where $G_{\mu\nu}^A$, $W_{\mu\nu}^A$, $B_{\mu\nu}$, and $B_{D\,\mu\nu}$ respectively represent the field strength tensors associated with the $SU(3)_C$, $SU(2)_L$, $U(1)_Y$, and $U(1)_D$ gauge groups, and $g_s$, $g$, $g'$, $g'_D$ denote the corresponding coupling constants. The dual field strength tensors are defined as $\tilde{V}_{\mu\nu} = \frac{1}{2} \epsilon_{\mu\nu\alpha\beta} V^{\alpha\beta}$ $(V = G^{A}$, $W^{A}$, $B$, $B_D)$ with $\epsilon^{0123} = 1$. The Wilson coefficients $C_{GG}$, $C_{WW}$, $C_{BB}$, and $C_{BB_D}$ parameterize the effective interactions induced by heavy degrees of freedom integrated out at higher energy scales.

Since the dark photon is the gauge boson of the abelian $U(1)_D$ symmetry, gauge invariance requires that its interaction before electroweak symmetry breaking is constructed solely from the hypercharge gauge field. Then, in this work we focus on the interaction described by the last term in Eq.~(\ref{eq:2.1}). After electroweak symmetry breaking, the corresponding dark axion portal interaction can be written as~\cite{Jodlowski:2024ayf}
\begin{eqnarray}
\begin{split}
\label{eq:2.2}
\mathcal{L} \supset
\frac{1}{2} g_{a\gamma\gamma'}\, a \, F^{\mu\nu}\tilde{F}_D^{\mu\nu}
+\frac{1}{2} g_{aZ\gamma'}\, a \, Z^{\mu\nu}\tilde{F}_D^{\mu\nu}.
\end{split}
\end{eqnarray}
The effective couplings are related by $g_{aZ\gamma'}=-\tan\theta_W\,g_{a\gamma\gamma'}$, which follows from electroweak gauge invariance, where $\theta_W$ is the Weinberg angle.

According to the above discussions, in the case of $m_a > m_{\gamma'}$, the ALP can decay through several channels, including the two-body modes $a \to \gamma \gamma'$ and $a \to Z \gamma'$, as well as the three-body decay into charged SM fermions $a \to \gamma' f\bar{f}$. The corresponding partial decay widths are given by ~\cite{Kaneta:2016wvf,Jodlowski:2023yne}
\begin{equation}
\label{eq:2.3}
\begin{aligned}
\Gamma_{a\to\gamma\gamma'} = \frac{g_{a\gamma\gamma'}^2}{32\pi} \, m_a^3 \left(1 - \frac{m_{\gamma'}^2}{m_a^2}\right)^3,
\end{aligned}
\end{equation}

\begin{equation}
\label{eq:2.4}
\begin{aligned}
\Gamma_{a\to Z\gamma'} = \frac{g_{aZ\gamma'}^{2}}{32\pi} \, m_a^{3} \left[ \left(1-\frac{m_Z^{2}+m_{\gamma'}^{2}}{m_a^{2}}\right)^{2}
-4\frac{m_Z^{2}m_{\gamma'}^{2}}{m_a^{4}}\right]^{3/2},
\end{aligned}
\end{equation}

\begin{equation}
\label{eq:2.5}
\begin{aligned}
\Gamma_{a\to\gamma'f\bar{f}} &\simeq Q_f^2 N_f \frac{\alpha_{\rm EM} g_{a\gamma\gamma'}^2}{192\pi^2 m_a^3} \Bigg[32 m_f^6 \coth^{-1}\!\left(\frac{m_a}{\sqrt{m_a^2 - 4m_f^2}}\right) \\
&\quad + m_a\sqrt{m_a^2 - 4m_f^2}\big(26 m_a^2 m_f^2 - 7m_a^4 + 8m_f^4\big) \\
&\quad - 4m_a^6 \log\left(\frac{2m_f}{\sqrt{m_a^2 - 4m_f^2}+m_a}\right) \\
&\quad + 12 m_a^2 m_f^4 \log\left(\frac{16 m_f^4 \big(m_a-\sqrt{m_a^2-4m_f^2}\big)}{\big(\sqrt{m_a^2-4m_f^2}+m_a\big)^5}\right) \Bigg],
\end{aligned}
\end{equation}
where $Q_f$ and $m_f$ denote the electromagnetic charge of the fermion $f$ in units of $e$ and its mass, respectively, $N_f$ is the color factor with $N_f=1~(3)$ for leptons (quarks), and $\alpha_{\rm EM}$ denotes the electromagnetic fine-structure constant. For ALP masses in the range $1~\mathrm{GeV}\leq m_a\leq1200~\mathrm{GeV}$ considered in this work, we assume $m_a\gg m_{\gamma'}$. When $m_a>m_Z$, the decay channel $a\to Z\gamma'$ becomes kinematically allowed, with its branching ratio increasing as the ALP mass increases and reaching approximately $20\%$ at $m_a=1200~\mathrm{GeV}$. In addition, for $m_a>2m_f$, the three-body decay $a\to\gamma' f\bar f$ is also kinematically allowed and its branching ratio is about $10\%$~\cite{Jodlowski:2024ayf}. Therefore, for the ALP with higher masses, despite the increasing branching ratio from the other decay channels, $a\to Z\gamma'$ and $a\to\gamma' f\bar f$, $a\to\gamma\gamma'$ remains the dominant decay mode and is taken as the signal decay channel in the following analysis.

The existing constraints on the $a$--$\gamma$--$\gamma'$ interaction mainly originate from cosmological, astrophysical, beam-dump and collider experiments. Cosmological and astrophysical observations, including those based on supernova cooling~\cite{Hook:2021ous}, CMB spectral distortions~\cite{Hook:2023smg}, and the effective number of relativistic species $\Delta N_{\rm eff}$~\cite{Hong:2023fcy}, provide stringent constraints on the dark axion portal in the light-mass region. These constraints primarily apply to ALPs with masses below the MeV scale and become particularly restrictive for ultralight ALPs with masses below the eV scale. Existing experimental bounds have been obtained from beam-dump experiments, such as CHARM~\cite{deNiverville:2018hrc,Gninenko:2011uv,Gninenko:2012eq} and NuCal~\cite{Jodlowski:2023yne,Blumlein:2011mv,Blumlein:1990ay}, which constrain the ALP parameter space for $m_a < 0.1~\mathrm{GeV}$. Additional constraints have also been obtained from electron--positron collider experiments. BaBar constrains the coupling down to $g_{a\gamma\gamma'}\sim2\times10^{-3}~\mathrm{GeV}^{-1}$ for $0.001~\mathrm{GeV}\lesssim m_a\lesssim10~\mathrm{GeV}$~\cite{deNiverville:2018hrc,BaBar:2008aby}. LEP reaches constraints at the level of $g_{a\gamma\gamma'}\sim10^{-4}~\mathrm{GeV}^{-1}$ over the mass range $0.001~\mathrm{GeV}\lesssim m_a\lesssim80~\mathrm{GeV}$~\cite{Jodlowski:2024ayf,OPAL:1994kgw,DELPHI:1996drf}, while LEP2 probes couplings down to $g_{a\gamma\gamma'}\sim10^{-3}~\mathrm{GeV}^{-1}$ for $0.01~\mathrm{GeV}\lesssim m_a\lesssim200~\mathrm{GeV}$~\cite{Jodlowski:2024ayf,DELPHI:2003dlq}.

\section{The possibility of detecting the ALP--photon--dark photon interaction at a muon collider}
\begin{figure}[H]
\begin{center}
\subfigure[]{\includegraphics [scale=0.5] {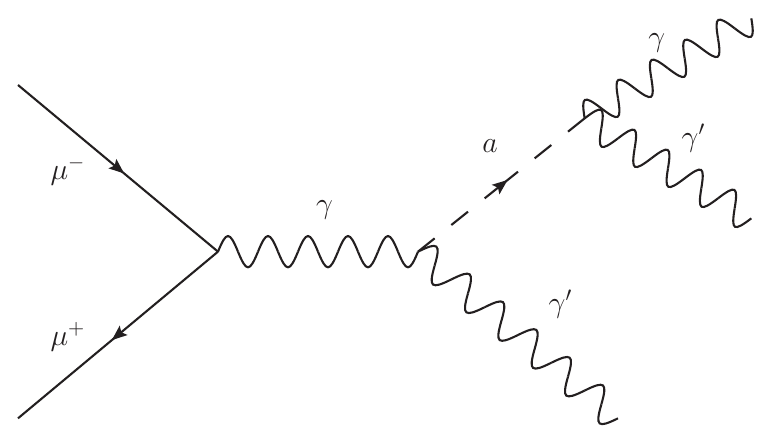}}
\subfigure[]{\includegraphics [scale=0.5] {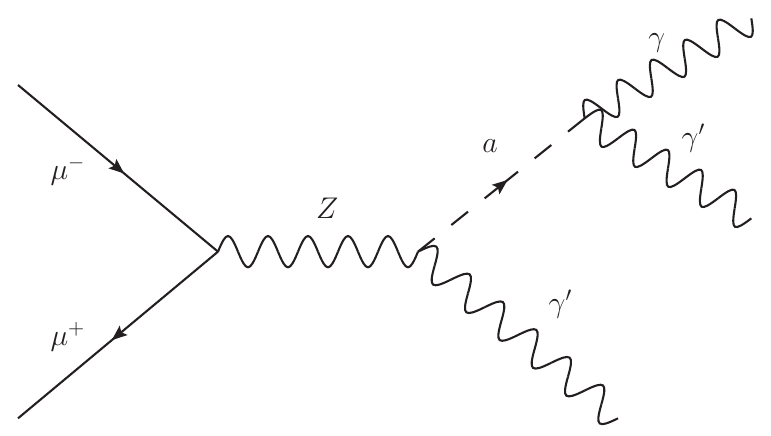}}
\caption{The Feynman diagrams for the signal process $\mu^+ \mu^- \to a \gamma'~(a \to \gamma \gamma')$ via $\gamma$ (a) and  $Z$ boson (b) exchanges.}
\label{fig:1}
\end{center}
\end{figure}

We consider the signal process $\mu^{+}\mu^{-} \to a \gamma'$ with $a \to \gamma \gamma'$ under the assumption that $m_a \gg m_{\gamma'}$, which proceeds via $s$-channel exchanges of a photon and a $Z$ boson, corresponding to diagrams~(a) and~(b) in Fig.~\ref{fig:1}. The subsequent decay of the ALP gives rise to a mono-photon signature accompanied by missing energy as the dark photons escape detection. The analysis is performed at the $1.5~\mathrm{TeV}$ muon collider with $\mathcal{L}=500~\mathrm{fb}^{-1}$.

We begin by implementing the effective Lagrangian in \verb"FeynRules"~\cite{Alloul:2013bka} to generate the corresponding UFO model file. Based on this model, signal events are generated at leading order using \verb"MadGraph5_aMC@NLO"~\cite{Alwall:2014hca} at the $1.5~\mathrm{TeV}$ muon collider with both polarized and unpolarized beam configurations. At the event generation level, basic cuts are imposed on the final-state photon to ensure detector acceptance.
\begin{eqnarray}
\begin{split}
\label{eq:2.6}
p_T^\gamma > 10~\mathrm{GeV}, \qquad |\eta^\gamma| < 2.5.
\end{split}
\end{eqnarray}
Here $p_T^\gamma$ denotes the transverse momentum of the photon and $|\eta^\gamma|$ is the absolute value of its pseudorapidity. These requirements ensure that the photon is sufficiently energetic and lies within the central region of the detector, where it can be efficiently reconstructed.

\begin{figure}[H]
\begin{center}
\includegraphics[scale=0.41]{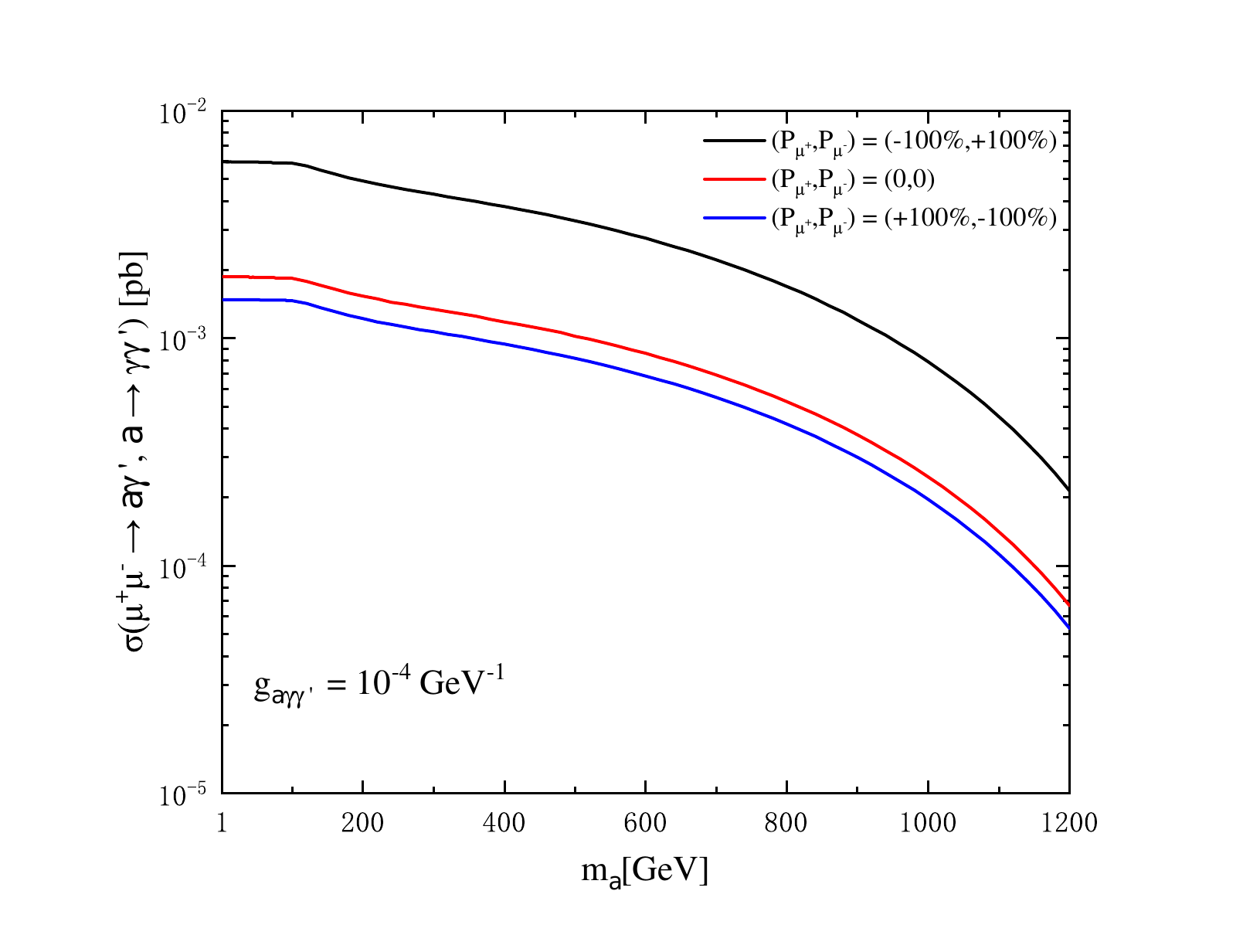}
\caption{Production cross sections for the signal process $\mu^+\mu^- \to a\gamma'~(a \to \gamma\gamma')$ as functions of the ALP mass $m_a$ at the $1.5~\mathrm{TeV}$  muon collider with the coupling $g_{a\gamma\gamma'} = 10^{-4}~\mathrm{GeV}^{-1}$ for different polarization options.}
\label{fig:2}
\end{center}
\end{figure}

After imposing the above basic cuts, we compute the production cross section of the signal process at the $1.5~\mathrm{TeV}$ muon collider. Fig.~\ref{fig:2} shows the dependence of the cross section on $m_a$ for $1~\mathrm{GeV} \leq m_a \leq 1200~\mathrm{GeV}$. We consider four beam polarization configurations, namely $P(\mu^+, \mu^-) = (+100\%, +100\%), (-100\%, -100\%), (+100\%, -100\%), (-100\%, +100\%)$, with a fixed effective coupling $g_{a\gamma\gamma'} = 10^{-4}~\mathrm{GeV}^{-1}$. The cross sections for $P(\mu^+, \mu^-) = (+100\%, +100\%)$ and $(-100\%, -100\%)$ are identical and highly suppressed at the level of $10^{-9}$ $\sim$ $10^{-10}~\mathrm{pb}$, and are therefore not shown in Fig.~\ref{fig:2}. As a result, Fig.~\ref{fig:2} displays three representative configurations, namely $P(\mu^+, \mu^-) = (-100\%, +100\%), (0, 0)$, and $(+100\%, -100\%)$, corresponding to the black, red, and blue curves, respectively. As the ALP mass increases, the production cross section gradually decreases. In the low-mass region $m_a \lesssim 1~\mathrm{GeV}$, the cross section is almost independent of $m_a$ due to $\sqrt{s}\gg m_a$. For $P(\mu^+, \mu^-) = (-100\%, +100\%)$, the values of the cross section decrease from $5.955 \times 10^{-3}$ to $2.139 \times 10^{-4}~\mathrm{pb}$ as $m_a$ increases from $1$ to $1200~\mathrm{GeV}$. For $P(\mu^+, \mu^-) = (+100\%, -100\%)$, the corresponding values vary from $1.459 \times 10^{-3}$ to $5.316 \times 10^{-5}~\mathrm{pb}$, while for the unpolarized case $P(\mu^+, \mu^-) = (0, 0)$, they decrease from $1.859 \times 10^{-3}$ to $6.675 \times 10^{-5}~\mathrm{pb}$.

To further investigate the kinematic features of the signal and evaluate the detection prospects, the generated parton-level events, including the effects of beam polarization, are subsequently passed to \verb"PYTHIA8"~\cite{Sjostrand:2014zea} for parton showering. Since the signal process does not involve colored final states, hadronization effects are negligible. Detector effects are simulated using \verb"DELPHES"~\cite{deFavereau:2013fsa} with a generic muon collider detector card. The reconstructed events are then analyzed with \verb"MadAnalysis5"~\cite{Conte:2012fm}, where kinematic distributions and cut analyses are evaluated.

\begin{figure}[H]
\begin{center}
\subfigure[]{\includegraphics [scale=0.3] {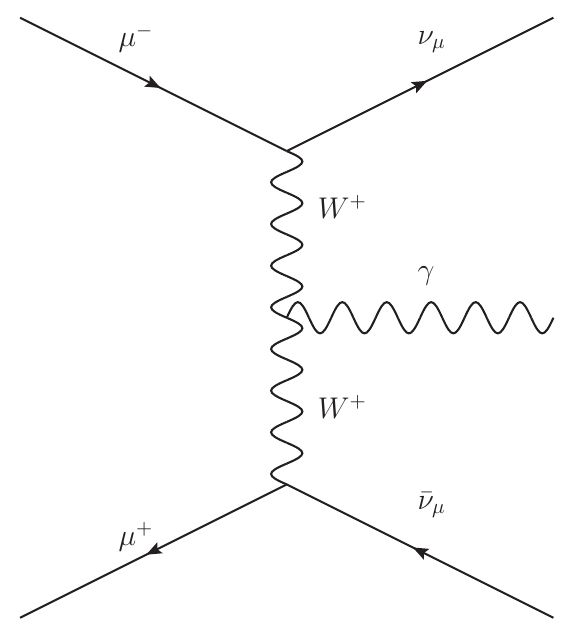}}
\subfigure[]{\includegraphics [scale=0.3] {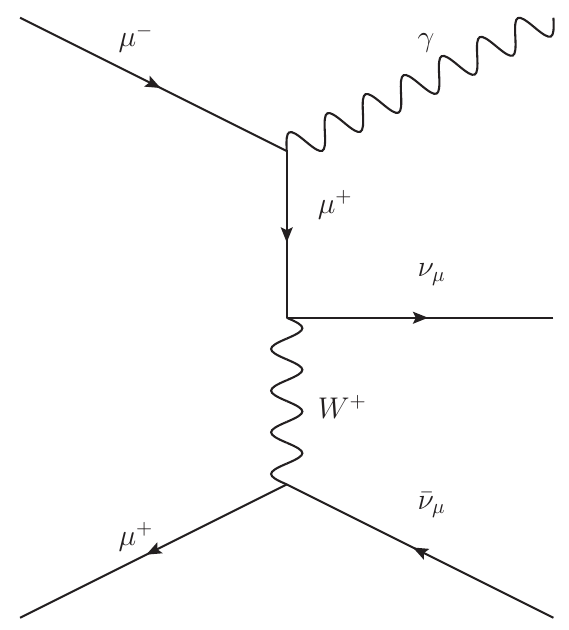}}
\subfigure[]{\includegraphics [scale=0.3] {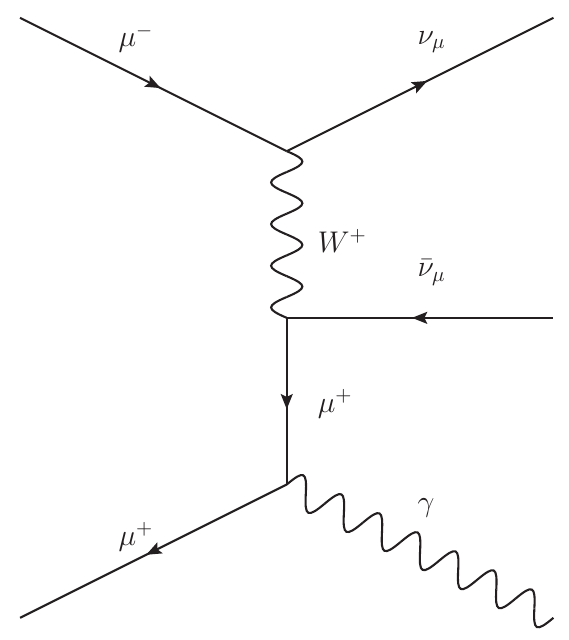}}
\subfigure[]{\includegraphics [scale=0.3] {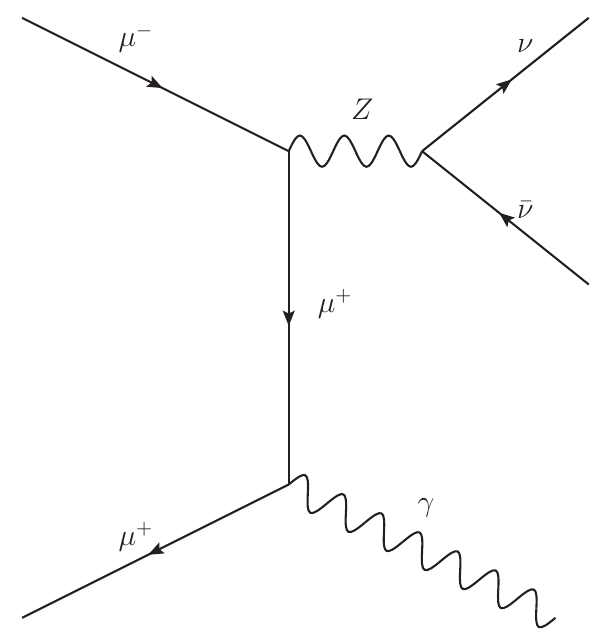}}
\subfigure[]{\includegraphics [scale=0.3] {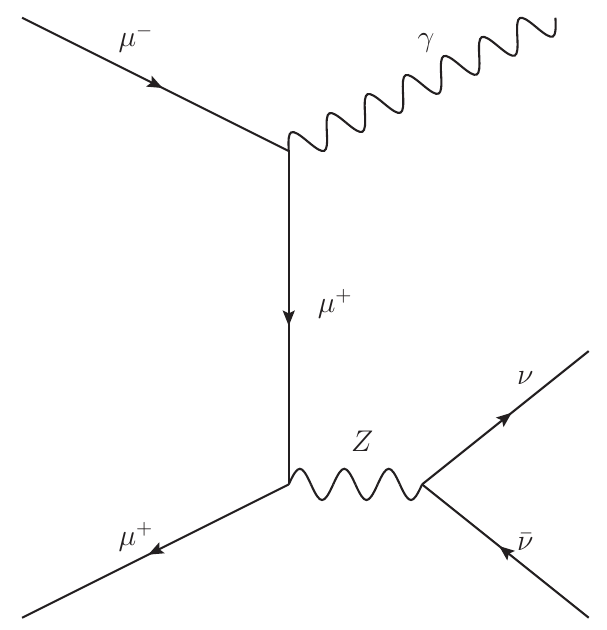}}
\caption{The typical Feynman diagrams for the SM background process $\mu^{+}\mu^{-}\to\nu\bar{\nu}\gamma$.}
\label{fig:3}
\end{center}
\end{figure}

For a realistic assessment of the signal observability, it is necessary to compare it with the corresponding SM background. The dominant contribution arises from the SM process $\mu^{+}\mu^{-} \to \nu \bar{\nu} \gamma$, which receives contributions from both $t$-channel $W$-boson exchange and $s$-channel $Z$-boson exchange with $Z \to \nu \bar{\nu}$ $(\nu=\nu_{e},\nu_\mu,\nu_\tau)$. The corresponding Feynman diagrams are presented separately in Fig.~\ref{fig:3}. In order to ensure a consistent comparison, the background is evaluated under the same beam polarization configuration as the signal. The values of the corresponding cross sections for the SM background at $\sqrt{s}=1.5~\mathrm{TeV}$ are $6.262 \times 10^{-2}~\mathrm{pb}$ for $P(\mu^+, \mu^-) = (-100\%, +100\%)$, $10.656~\mathrm{pb}$ for $P(\mu^+, \mu^-) = (+100\%, -100\%)$, and $2.680~\mathrm{pb}$ for the unpolarized case of $P(\mu^+, \mu^-) = (0,0)$. Among the polarized beam configurations, $P(\mu^+, \mu^-) = (-100\%, +100\%)$ corresponds to the largest signal cross section and the smallest background cross section. Therefore, this configuration is adopted as the benchmark polarized scenario. However, such a fully polarized beam configuration represents an idealized scenario and may be challenging to realize in practice. To account for this limitation, the unpolarized configuration $P(\mu^+, \mu^-) = (0,0)$ is also considered in the following analysis for comparison.

\begin{figure}[H]
\centering

\begin{minipage}[b]{0.45\textwidth}
\centering
\includegraphics[width=0.95\textwidth,trim=8 5 5 5,clip]{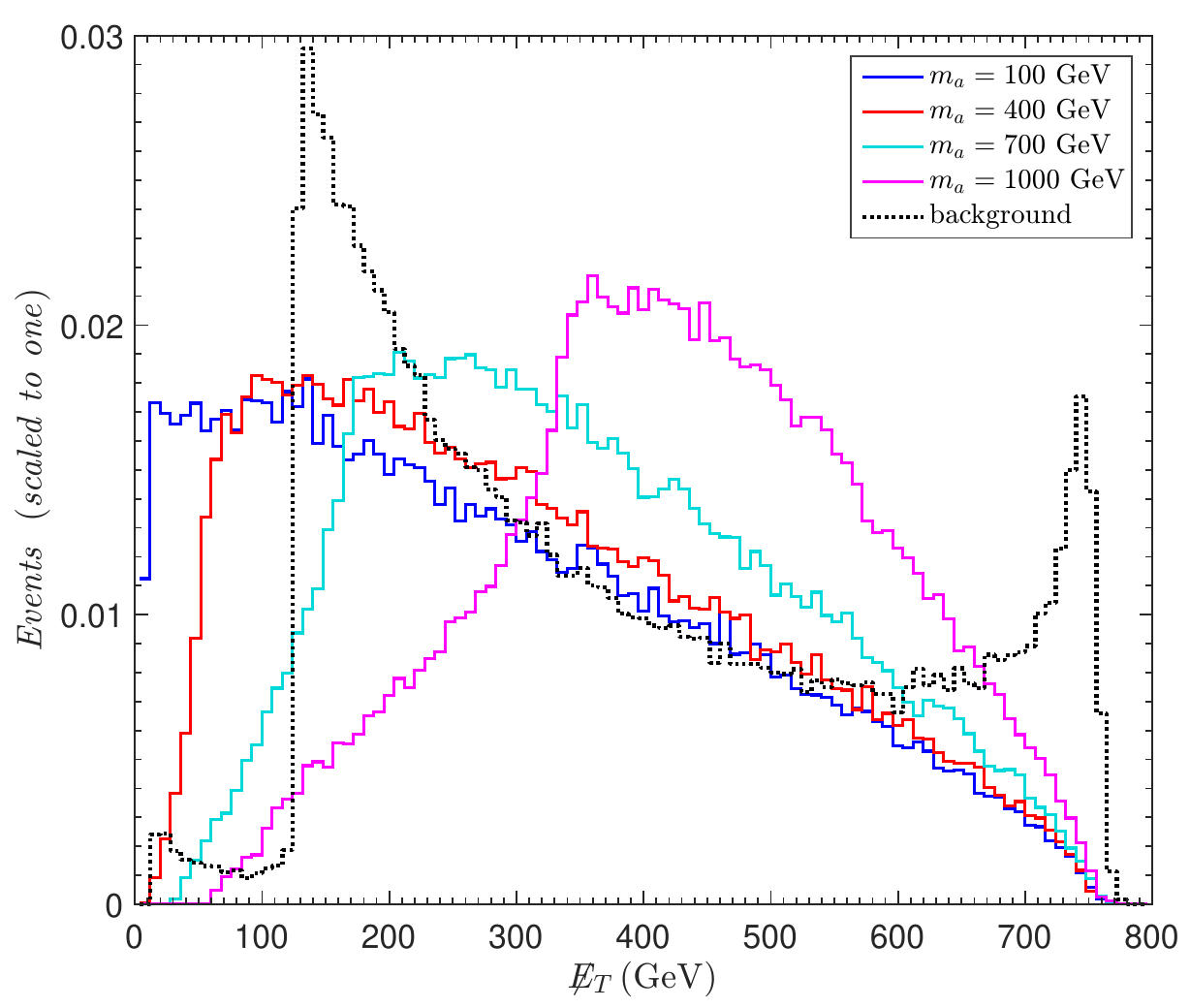}\\[-2mm]
(a)
\end{minipage}
\hspace{0.025\textwidth}
\begin{minipage}[b]{0.45\textwidth}
\centering
\includegraphics[width=0.95\textwidth,trim=8 5 5 5,clip]{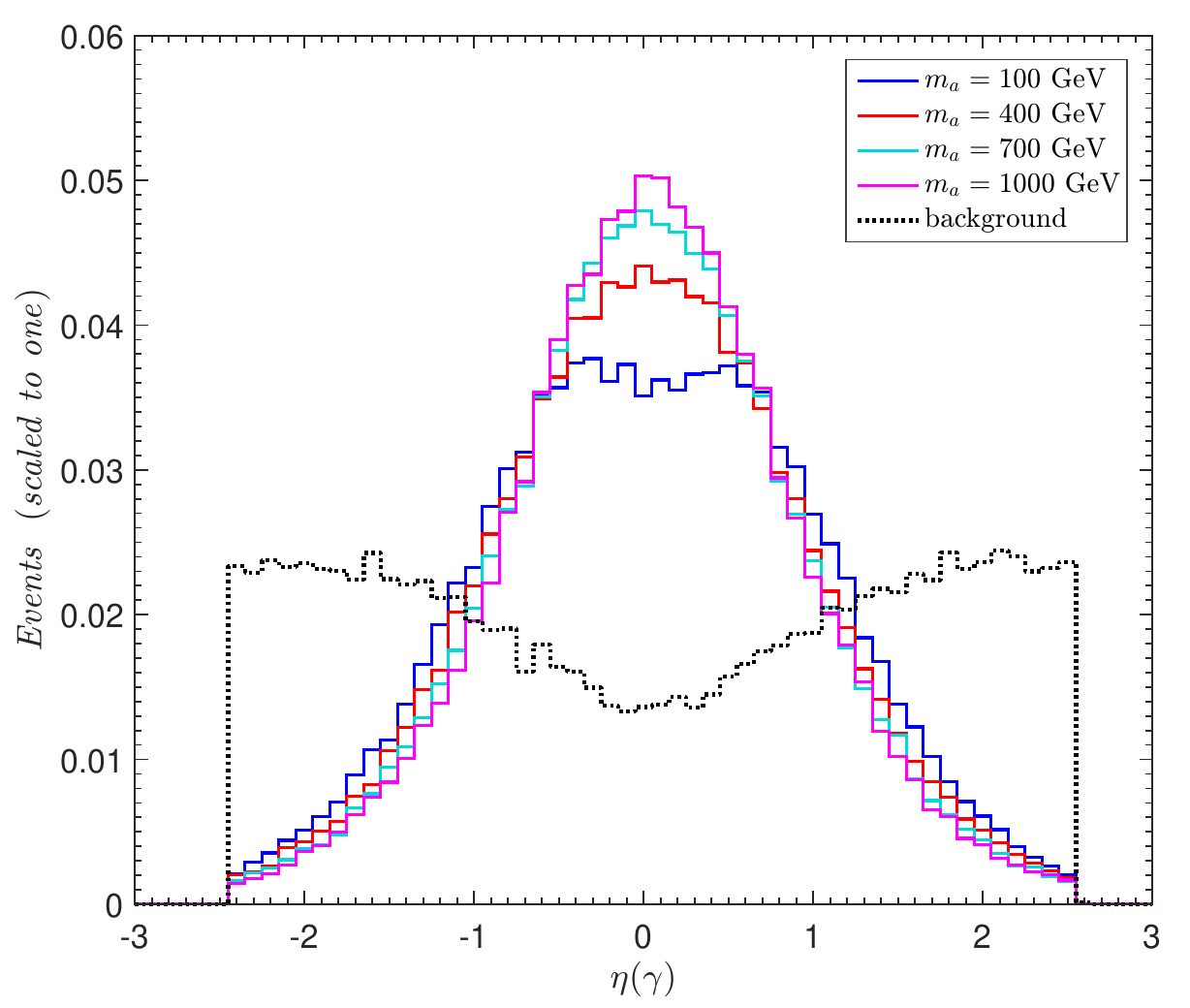}\\[-2mm]
(b)
\end{minipage}

\vspace{0.3cm}

\begin{minipage}[b]{0.45\textwidth}
\centering
\includegraphics[width=0.95\textwidth,trim=8 5 5 5,clip]{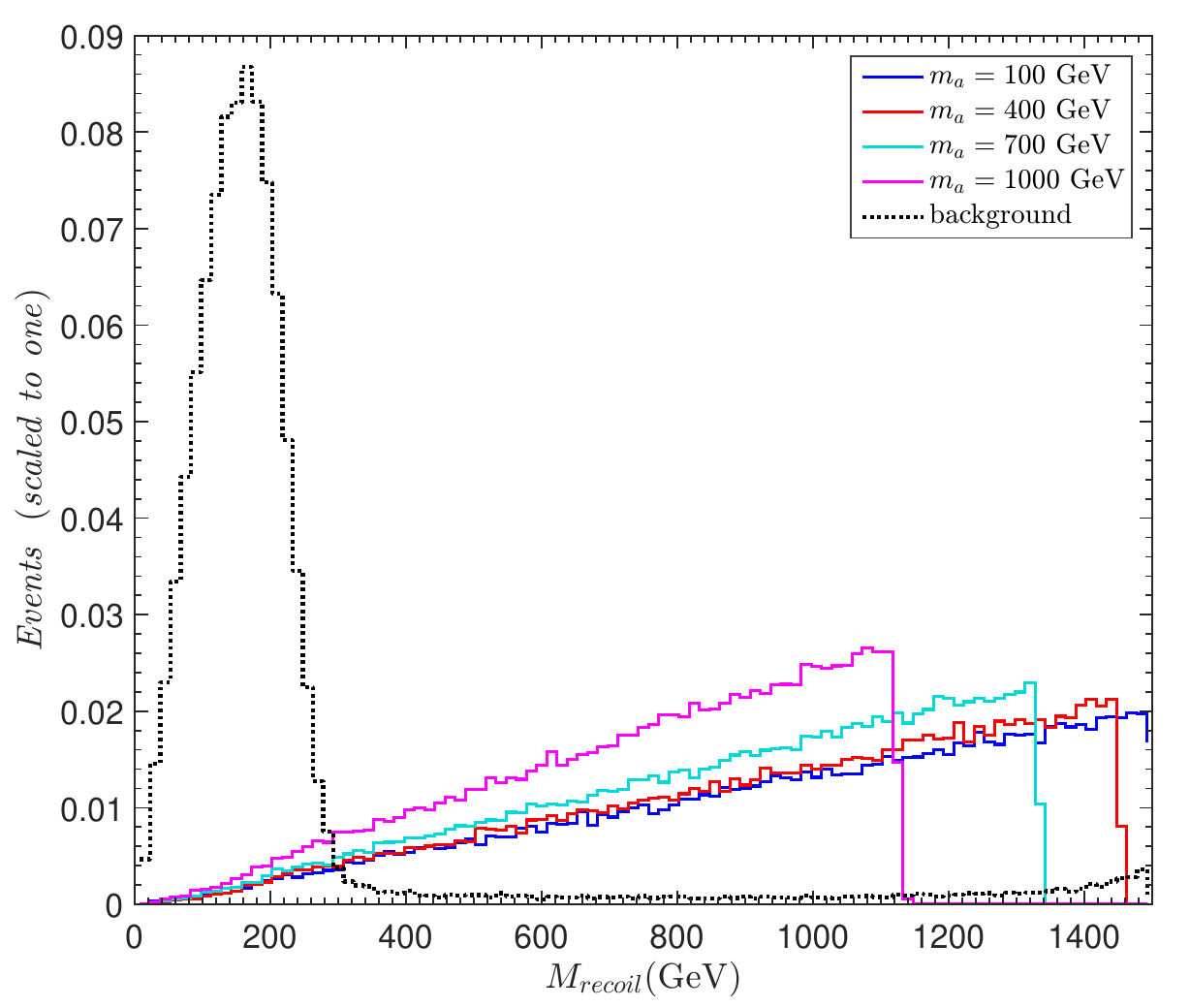}\\[-2mm]
(c)
\end{minipage}
\hspace{0.025\textwidth}
\begin{minipage}[b]{0.45\textwidth}
\centering
\includegraphics[width=0.95\textwidth,trim=8 5 5 5,clip]{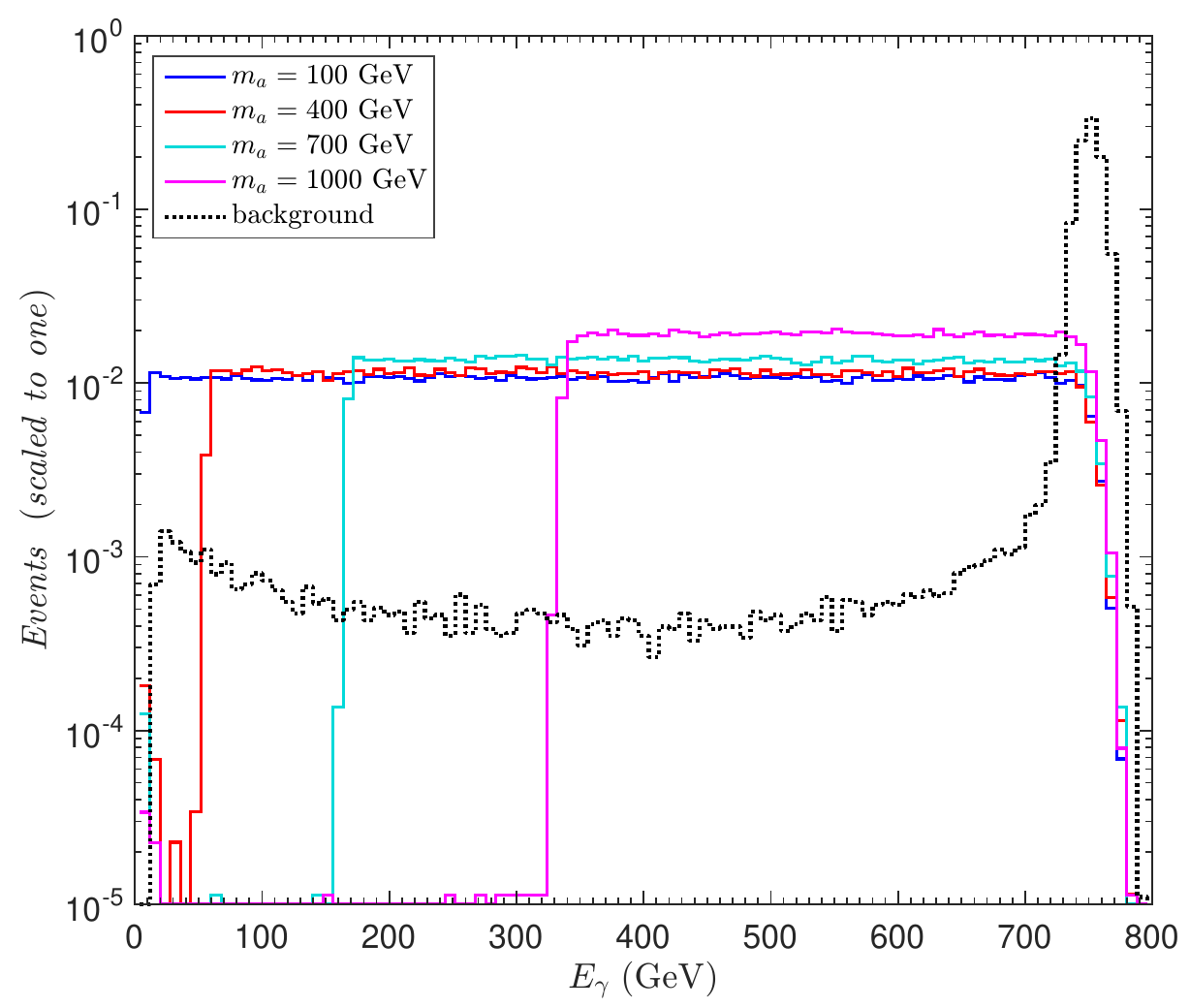}\\[-2mm]
(d)
\end{minipage}

\caption{The normalized distributions of the observables $\slashed{E}_T$ (a), $\eta_{\gamma}$ (b), $M_{\rm recoil}$ (c), and $E_{\gamma}$ (d) for the signal at selected ALP-mass benchmark points and the SM background at the $1.5~\mathrm{TeV}$ muon collider with $\mathcal{L}=500~\mathrm{fb}^{-1}$ and $P(\mu^+,\mu^-)=(-100\%,+100\%)$.}
\label{fig:4}
\end{figure}

\begin{figure}[H]
\centering

\begin{minipage}[b]{0.45\textwidth}
\centering
\includegraphics[width=0.95\textwidth,trim=8 5 5 5,clip]{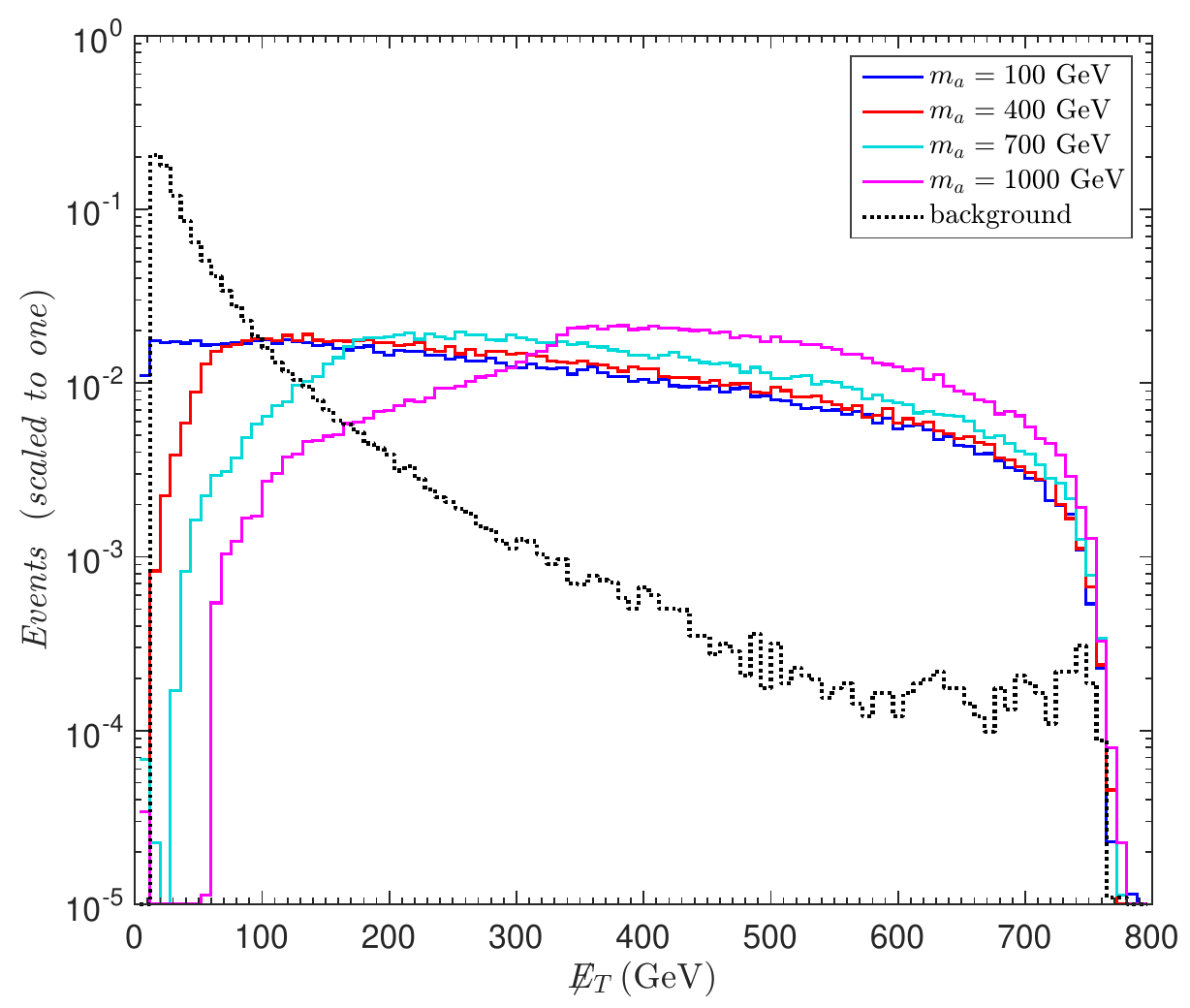}\\[-2mm]
(a)
\end{minipage}
\hspace{0.025\textwidth}
\begin{minipage}[b]{0.45\textwidth}
\centering
\includegraphics[width=0.95\textwidth,trim=8 5 5 5,clip]{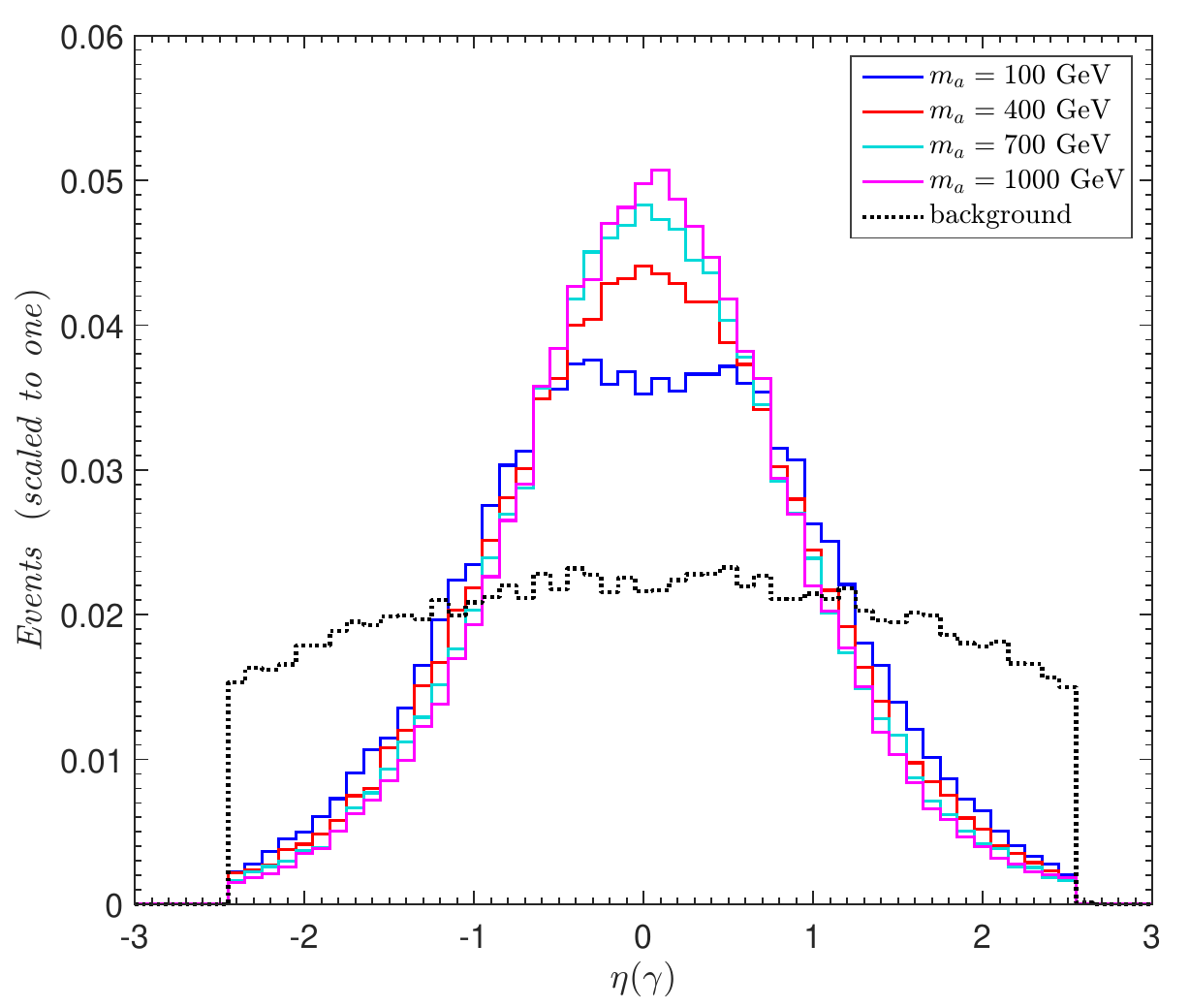}\\[-2mm]
(b)
\end{minipage}

\vspace{0.3cm}

\begin{minipage}[b]{0.45\textwidth}
\centering
\includegraphics[width=0.95\textwidth,trim=8 5 5 5,clip]{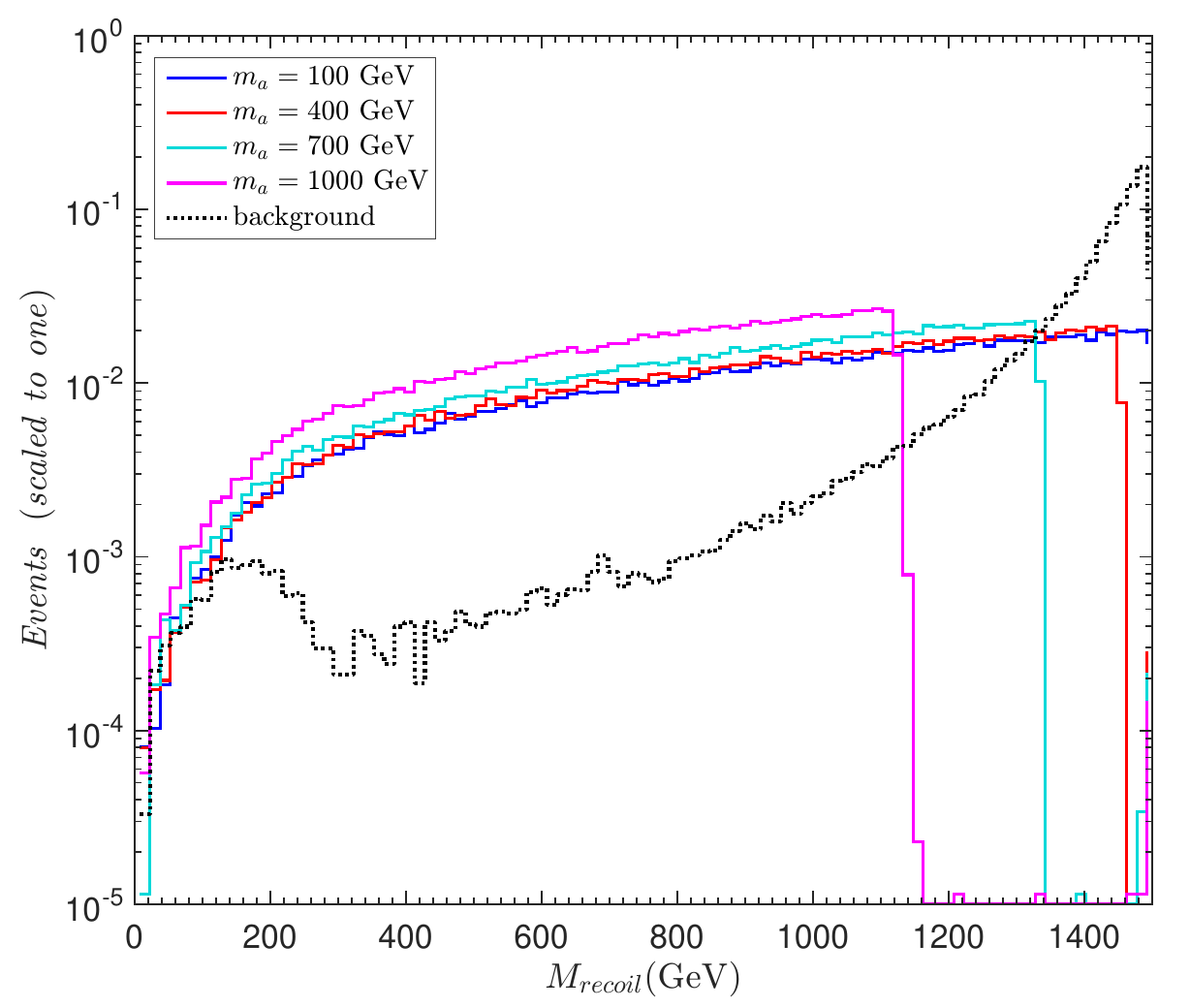}\\[-2mm]
(c)
\end{minipage}
\hspace{0.025\textwidth}
\begin{minipage}[b]{0.45\textwidth}
\centering
\includegraphics[width=0.95\textwidth,trim=8 5 5 5,clip]{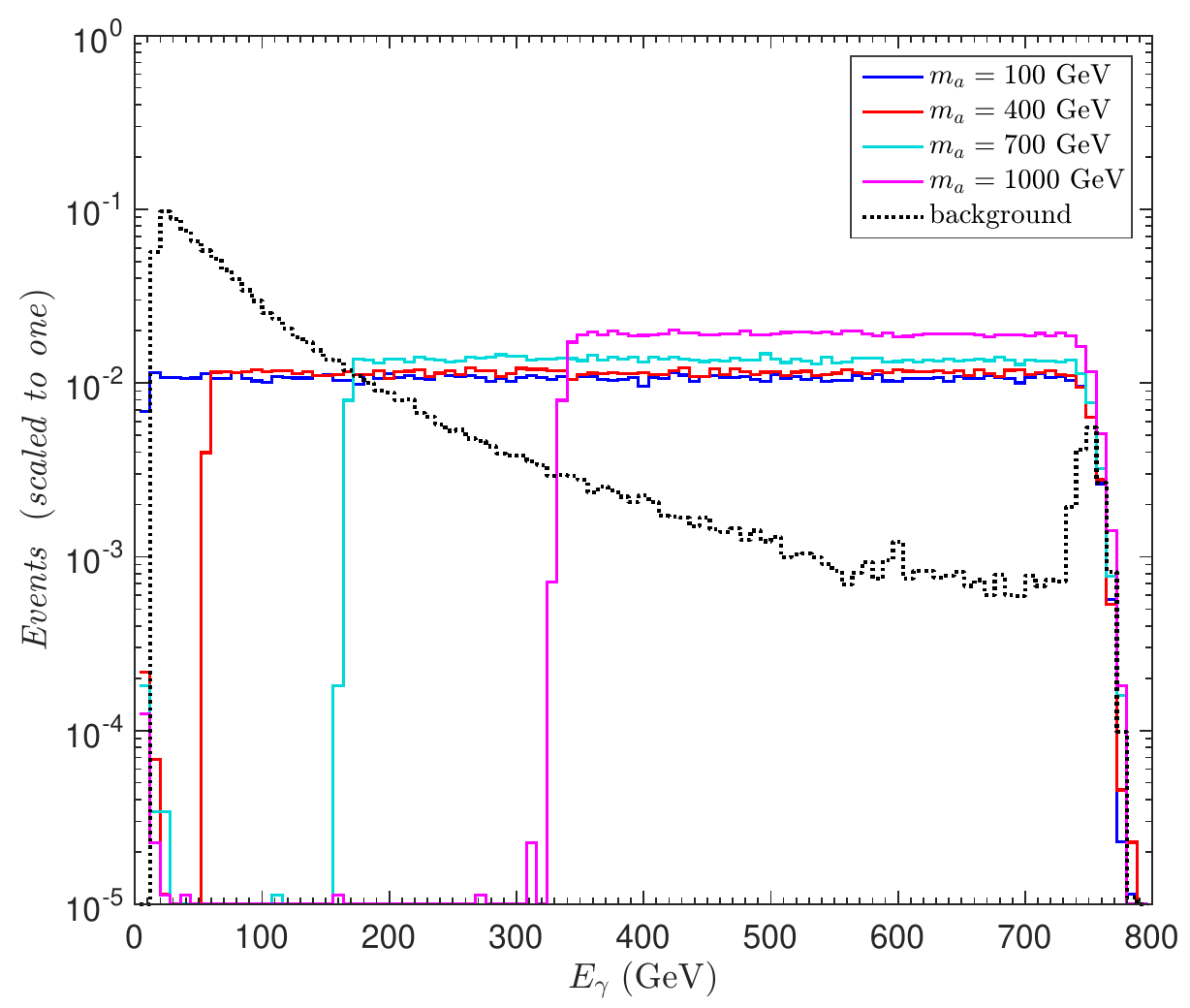}\\[-2mm]
(d)
\end{minipage}

\caption{Same as Fig.~\ref{fig:4} but for the unpolarized beam configuration $P(\mu^+,\mu^-)=(0,0)$.}
\label{fig:5}
\end{figure}

Several kinematic observables are used to distinguish the signal from the SM background, including $\slashed{E}_T$, $\eta_{\gamma}$, $M_{\rm recoil}$, and $E_{\gamma}$, where $\slashed{E}_T$ corresponds to the missing transverse energy arising from the undetected dark photons in the final state, $\eta_{\gamma}$ denotes the pseudorapidity of the final-state photon, $M_{\rm recoil}$ is defined as the recoil mass of the system recoiling against the detected photon, and $E_{\gamma}$ represents the photon energy. The normalized distributions of these observables for the signal and SM background are shown in Figs.~\ref{fig:4} and~\ref{fig:5} for the representative benchmark points $m_a=100$, 400, 700, and $1000~\mathrm{GeV}$ at the $1.5~\mathrm{TeV}$ muon collider with $\mathcal{L}=500~\mathrm{fb}^{-1}$ for $P(\mu^+,\mu^-)=(-100\%,+100\%)$ and $P(\mu^+,\mu^-)=(0,0)$, respectively.

For the polarized beam configuration, Fig.~\ref{fig:4}(a) presents the signal distributions with peaks at different values of $\slashed{E}_T$, which shift toward higher $\slashed{E}_T$ as the ALP mass increases, whereas the SM background features two pronounced peaks around $\slashed{E}_T\approx140~\mathrm{GeV}$ and $\slashed{E}_T\approx750~\mathrm{GeV}$. The lower-$\slashed{E}_T$ peak is mainly associated with the $t$-channel $W$-exchange contribution, while the peak near the kinematic endpoint is dominated by the radiative-return process $\mu^{+}\mu^{-}\to Z\gamma$, followed by $Z\to\nu\bar{\nu}$, yielding a neutrino pair accompanied by a hard photon. Between the two peaks, the background drops to a relatively low level. For $\slashed{E}_T\lesssim100~\mathrm{GeV}$, the background is strongly suppressed because the radiative-return contribution is absent in this region, while the remaining non-resonant electroweak contributions are relatively small. As shown in Fig.~\ref{fig:4}(b), the signal events are concentrated in the central region, with the $\eta_{\gamma}$ distribution peaking around $\eta_{\gamma}\approx0$. In contrast, the SM background is suppressed in the central region and exhibits a broader distribution toward larger $|\eta_{\gamma}|$. Figure~\ref{fig:4}(c) displays the recoil-mass distributions, where the signal distributions extend toward larger $M_{\rm recoil}$ before reaching their mass-dependent kinematic endpoints. In contrast, the SM background exhibits a pronounced peak around $M_{\rm recoil}\approx170~\mathrm{GeV}$ and then decreases rapidly with increasing $M_{\rm recoil}$. Finally, Fig.~\ref{fig:4}(d) presents the $E_{\gamma}$ distributions, where the signal distributions are approximately flat over most of their kinematically allowed energy ranges, while their lower endpoints shift toward higher $E_{\gamma}$ as the ALP mass increases. In contrast, the SM background exhibits a pronounced peak near the kinematic endpoint, $E_{\gamma}\approx750~\mathrm{GeV}$. This peak arises from the radiative-return process $\mu^{+}\mu^{-}\to Z\gamma$, in which the emitted photon carries nearly half of the center-of-mass energy, causing the $Z$ boson to be produced nearly on shell before decaying into a neutrino pair. Away from the kinematic endpoint, the radiative-return contribution is no longer enhanced, and the background decreases rapidly.

For comparison, the corresponding distributions for the unpolarized beam configuration are presented in Fig.~\ref{fig:5}. The signal distributions exhibit overall kinematic features similar to those in the polarized case, while noticeable differences appear in the SM background. In particular, the background is strongly enhanced at low $\slashed{E}_T$ and low $E_\gamma$, whereas the suppression of the SM background in the central region is weaker, leading to a relatively flat distribution around $\eta_\gamma\approx0$. For $M_{\rm recoil}$, the background decreases at low recoil masses and subsequently rises toward the high-$M_{\rm recoil}$ region.

Based on the characteristic features of these distributions, the optimized selection cuts for enhancing the signal and suppressing the SM background are summarized in Table~\ref{tab1}. These cuts are designed to select central, high-energy photon events with large missing transverse energy, which are characteristic of the signal topology. After applying all optimized cuts, the SM background is significantly suppressed, while a sizable fraction of the signal is retained. The signal and corresponding background production cross sections after the step-by-step optimized cuts are presented in Tables~\ref{tab2} and~\ref{tab3} for representative benchmark points $m_a = 100$, 400, 700, and $1000~\mathrm{GeV}$ with the effective coupling fixed at $g_{a\gamma\gamma'} = 10^{-4}~\mathrm{GeV}^{-1}$. Table~\ref{tab2} summarizes the results for the beam polarizations $P(\mu^+,\mu^-)=(-100\%,+100\%)$, while Table~\ref{tab3} presents the corresponding results for the unpolarized beam configuration $P(\mu^+,\mu^-)=(0,0)$. In addition, the statistical significance for each benchmark point is also shown in Tables~\ref{tab2} and~\ref{tab3}. The statistical significance~($SS$) is evaluated using $SS = S/\sqrt{S+B}$, where $S$ and $B$ denote the expected numbers of signal and background events obtained for the integrated luminosity $\mathcal{L}=500~\mathrm{fb}^{-1}$. The resulting $SS$ values range from $35.965$ to $8.105$ for the polarized case and from $1.836$ to $0.349$ for the unpolarized case.

\begin{table}[H]
\begin{center}
\setlength{\tabcolsep}{1.5mm}{
\caption{The optimized cuts on the signal and background for $1~\mathrm{GeV} \leq m_a \leq 1200~\mathrm{GeV}$ at the $1.5~\mathrm{TeV}$ muon collider with $\mathcal{L}=500~\mathrm{fb}^{-1}$.}
\label{tab1}
\resizebox{14cm}{!}{
\begin{tabular}[c]{l|c c}
\hline \hline
\multirow{2}{*}{Cuts}
& \multicolumn{2}{c}{$1~\mathrm{GeV} \leq m_a \leq 1200~\mathrm{GeV}$} \\
\cline{2-3}
& $P(\mu^+,\mu^-)=(-100\%,+100\%)$
& $P(\mu^+,\mu^-)=(0,0)$ \\
\hline
Cut 1: the photon number in the final state
& $N_{\gamma}\geq 1$
& $N_{\gamma}\geq 1$ \\
Cut 2: missing transverse energy
& $10~\mathrm{GeV}<\slashed{E}_T<670~\mathrm{GeV}$
& $\slashed{E}_T>100~\mathrm{GeV}$ \\
Cut 3: photon pseudorapidity
& $|\eta_{\gamma}|<1.2$
& $|\eta_{\gamma}|<1.2$ \\
Cut 4: recoil mass
& $M_{\rm recoil}>300~\mathrm{GeV}$
& $120~\mathrm{GeV}<M_{\rm recoil}<1320~\mathrm{GeV}$ \\
\hline \hline
\end{tabular}}}
\end{center}
\end{table}

\begin{table}[H]
\centering
\scriptsize
\setlength{\tabcolsep}{2.5mm}
\caption{The production cross sections of the signal and SM background for representative ALP mass points with $g_{a\gamma\gamma'} = 10^{-4}~\mathrm{GeV}^{-1}$ after the step-by-step optimized cuts employed at the $1.5~\mathrm{TeV}$ muon collider with $\mathcal{L}=500~\mathrm{fb}^{-1}$ and $P(\mu^+,\mu^-) = (-100\%,+100\%)$.}
\label{tab2}
\begin{tabular}{c|c c c c}
\hline \hline
\multirow{2}{*}{Cuts} & \multicolumn{4}{c}{Cross sections for signal (background) [pb]} \\
\cline{2-5}
 & $m_a=100$ GeV & $m_a=400$ GeV & $m_a=700$ GeV & $m_a=1000$ GeV \\
\hline
Basic Cuts
& \makecell{$5.867\times10^{-3}$\\($6.262\times10^{-2}$)}
& \makecell{$3.789\times10^{-3}$\\($6.262\times10^{-2}$)}
& \makecell{$2.211\times10^{-3}$\\($6.262\times10^{-2}$)}
& \makecell{$7.871\times10^{-4}$\\($6.262\times10^{-2}$)} \\
Cut 1
& \makecell{$5.130\times10^{-3}$\\($5.688\times10^{-2}$)}
& \makecell{$3.335\times10^{-3}$\\($5.688\times10^{-2}$)}
& \makecell{$1.949\times10^{-3}$\\($5.688\times10^{-2}$)}
& \makecell{$6.978\times10^{-4}$\\($5.688\times10^{-2}$)} \\
Cut 2
& \makecell{$4.927\times10^{-3}$\\($4.982\times10^{-2}$)}
& \makecell{$3.248\times10^{-3}$\\($4.982\times10^{-2}$)}
& \makecell{$1.886\times10^{-3}$\\($4.982\times10^{-2}$)}
& \makecell{$6.656\times10^{-4}$\\($4.982\times10^{-2}$)} \\
Cut 3
& \makecell{$3.814\times10^{-3}$\\($1.585\times10^{-2}$)}
& \makecell{$2.618\times10^{-3}$\\($1.585\times10^{-2}$)}
& \makecell{$1.559\times10^{-3}$\\($1.585\times10^{-2}$)}
& \makecell{$5.553\times10^{-4}$\\($1.585\times10^{-2}$)} \\
Cut 4
& \makecell{$3.717\times10^{-3}$\\($1.624\times10^{-3}$)}
& \makecell{$2.555\times10^{-3}$\\($1.624\times10^{-3}$)}
& \makecell{$1.516\times10^{-3}$\\($1.624\times10^{-3}$)}
& \makecell{$5.322\times10^{-4}$\\($1.624\times10^{-3}$)} \\
\hline
$SS$  & $35.965$ & $27.946$ & $19.128$ & $8.105$ \\
\hline \hline
\end{tabular}
\end{table}

\begin{table}[H]
\centering
\scriptsize
\setlength{\tabcolsep}{2.5mm}
\caption{Same as Table~\ref{tab2} but for the unpolarized beam configuration $P(\mu^+,\mu^-) = (0,0)$.}
\label{tab3}
\begin{tabular}{c|c c c c}
\hline \hline
\multirow{2}{*}{Cuts} & \multicolumn{4}{c}{Cross sections for signal (background) [pb]} \\
\cline{2-5}
 & $m_a=100$ GeV & $m_a=400$ GeV & $m_a=700$ GeV & $m_a=1000$ GeV \\
\hline
Basic Cuts
& \makecell{$1.831\times10^{-3}$\\($2.680$)}
& \makecell{$1.183\times10^{-3}$\\($2.680$)}
& \makecell{$6.901\times10^{-4}$\\($2.680$)}
& \makecell{$2.457\times10^{-4}$\\($2.680$)} \\
Cut 1
& \makecell{$1.606\times10^{-3}$\\($2.446$)}
& \makecell{$1.041\times10^{-3}$\\($2.446$)}
& \makecell{$6.088\times10^{-4}$\\($2.446$)}
& \makecell{$2.173\times10^{-4}$\\($2.446$)} \\
Cut 2
& \makecell{$1.272\times10^{-3}$\\($3.495\times10^{-1}$)}
& \makecell{$9.082\times10^{-4}$\\($3.495\times10^{-1}$)}
& \makecell{$5.915\times10^{-4}$\\($3.495\times10^{-1}$)}
& \makecell{$2.157\times10^{-4}$\\($3.495\times10^{-1}$)} \\
Cut 3
& \makecell{$1.041\times10^{-3}$\\($2.099\times10^{-1}$)}
& \makecell{$7.602\times10^{-4}$\\($2.099\times10^{-1}$)}
& \makecell{$5.060\times10^{-4}$\\($2.099\times10^{-1}$)}
& \makecell{$1.832\times10^{-4}$\\($2.099\times10^{-1}$)} \\
Cut 4
& \makecell{$9.519\times10^{-4}$\\($1.335\times10^{-1}$)}
& \makecell{$6.856\times10^{-4}$\\($1.335\times10^{-1}$)}
& \makecell{$4.965\times10^{-4}$\\($1.335\times10^{-1}$)}
& \makecell{$1.804\times10^{-4}$\\($1.335\times10^{-1}$)} \\
\hline
$SS$  & $1.836$ & $1.323$ & $0.959$ & $0.349$ \\
\hline \hline
\end{tabular}
\end{table}

\begin{figure}[H]
\begin{center}
\subfigure[]{\includegraphics [scale=0.28] {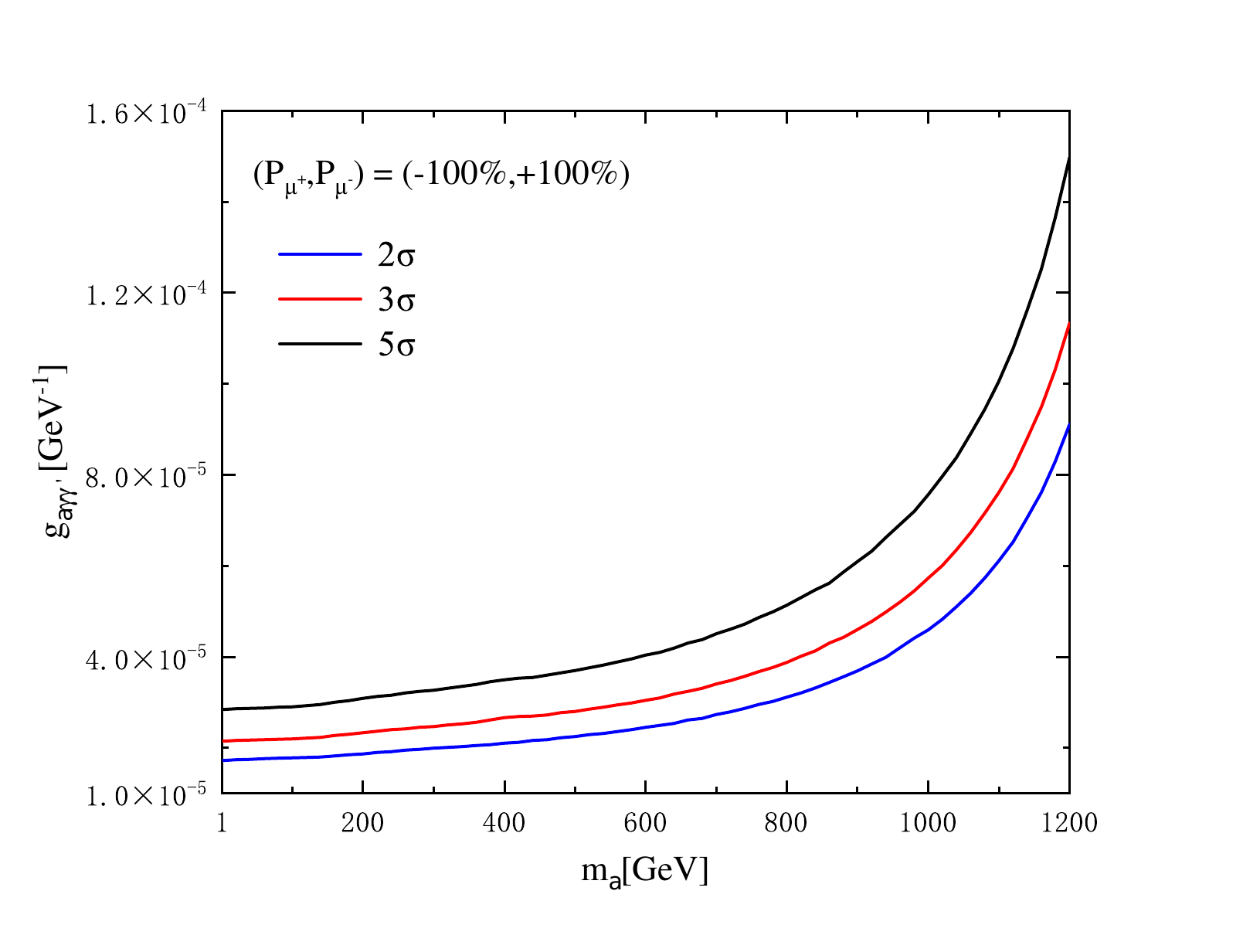}}
\subfigure[]{\includegraphics [scale=0.28] {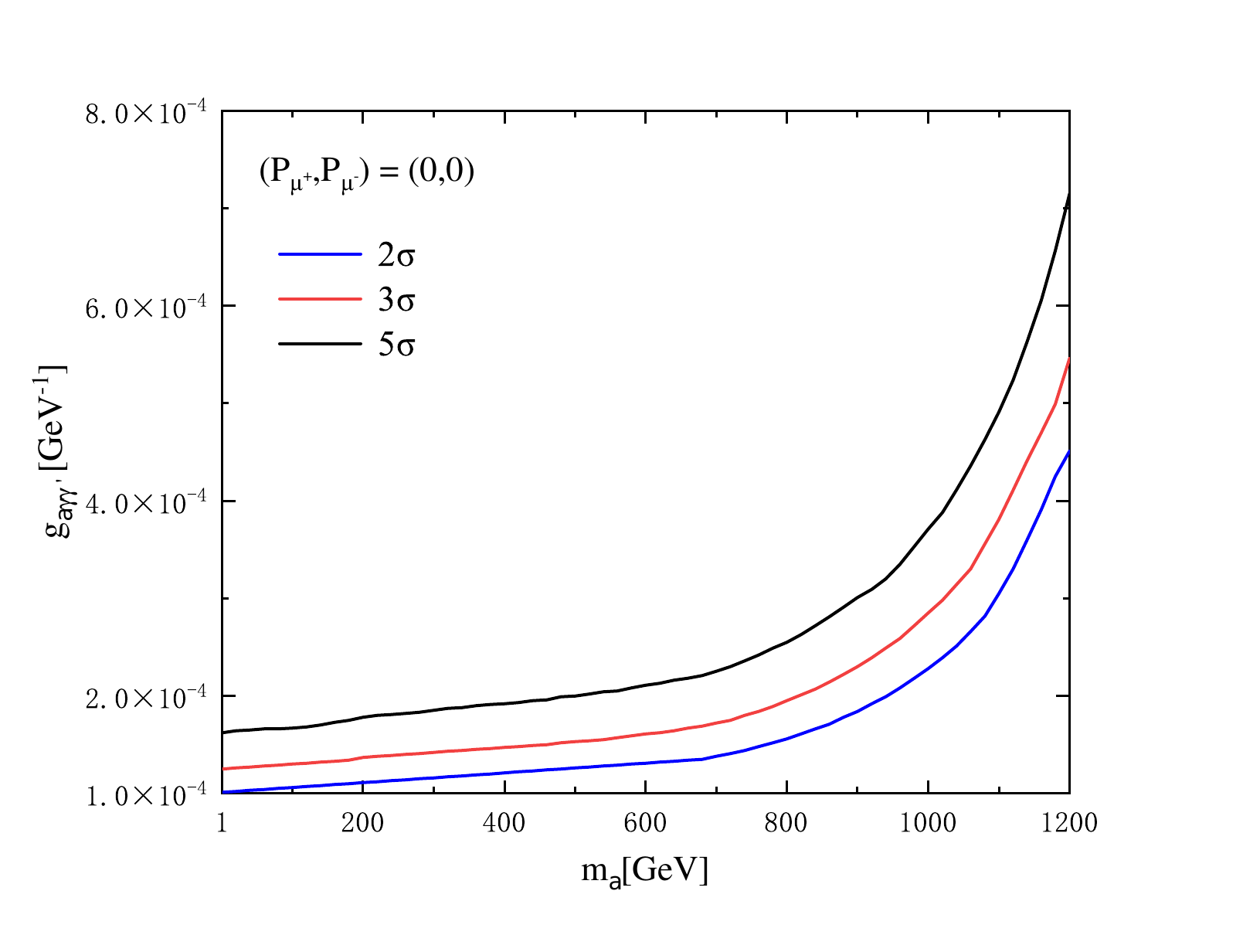}}
\caption{The projected $2\sigma$, $3\sigma$, and $5\sigma$ sensitivity curves in the $m_a$--$g_{a\gamma\gamma'}$ plane for the mono-photon final-state process $\mu^{+}\mu^{-}\to a\gamma'~(a\to\gamma\gamma')$ at the $1.5~\mathrm{TeV}$ muon collider with $\mathcal{L}=500~\mathrm{fb}^{-1}$ for both $P(\mu^+,\mu^-)=(-100\%,+100\%)$ (a) and $P(\mu^+,\mu^-)=(0,0)$ (b).}
\label{fig:6}
\end{center}
\end{figure}

In Fig.~\ref{fig:6}, we present the projected $2\sigma$, $3\sigma$, and $5\sigma$ sensitivity curves in the $m_a-g_{a\gamma\gamma'}$ plane for both the beam polarizations $P(\mu^+,\mu^-)=(-100\%,+100\%)$ and the unpolarized beam configuration $P(\mu^+,\mu^-)=(0,0)$ at the $1.5~\mathrm{TeV}$ muon collider with $\mathcal{L}=500~\mathrm{fb}^{-1}$. As shown in Fig.~\ref{fig:6}, the sensitivities gradually decrease as the ALP mass increases. For ALP masses satisfying $1~\mathrm{GeV} \leq m_a \leq 1200~\mathrm{GeV}$, the prospective sensitivities to $g_{a\gamma\gamma'}$ can be obtained as $1.720\times10^{-5}~(1.010\times10^{-4})$ $\sim$ $9.110\times10^{-5}~(4.510\times10^{-4})~\mathrm{GeV}^{-1}$, $2.150\times10^{-5}~(1.250\times10^{-4})$ $\sim$ $1.133\times10^{-4}~(5.460\times10^{-4})~\mathrm{GeV}^{-1}$, and $2.840\times10^{-5}~(1.620\times10^{-4})$ $\sim$ $1.497\times10^{-4}~(7.150\times10^{-4})~\mathrm{GeV}^{-1}$ at the $2\sigma$, $3\sigma$, and $5\sigma$ confidence levels for the polarized (unpolarized) beam configuration, respectively.

\begin{figure}[H]
\centering
\includegraphics[scale=0.41]{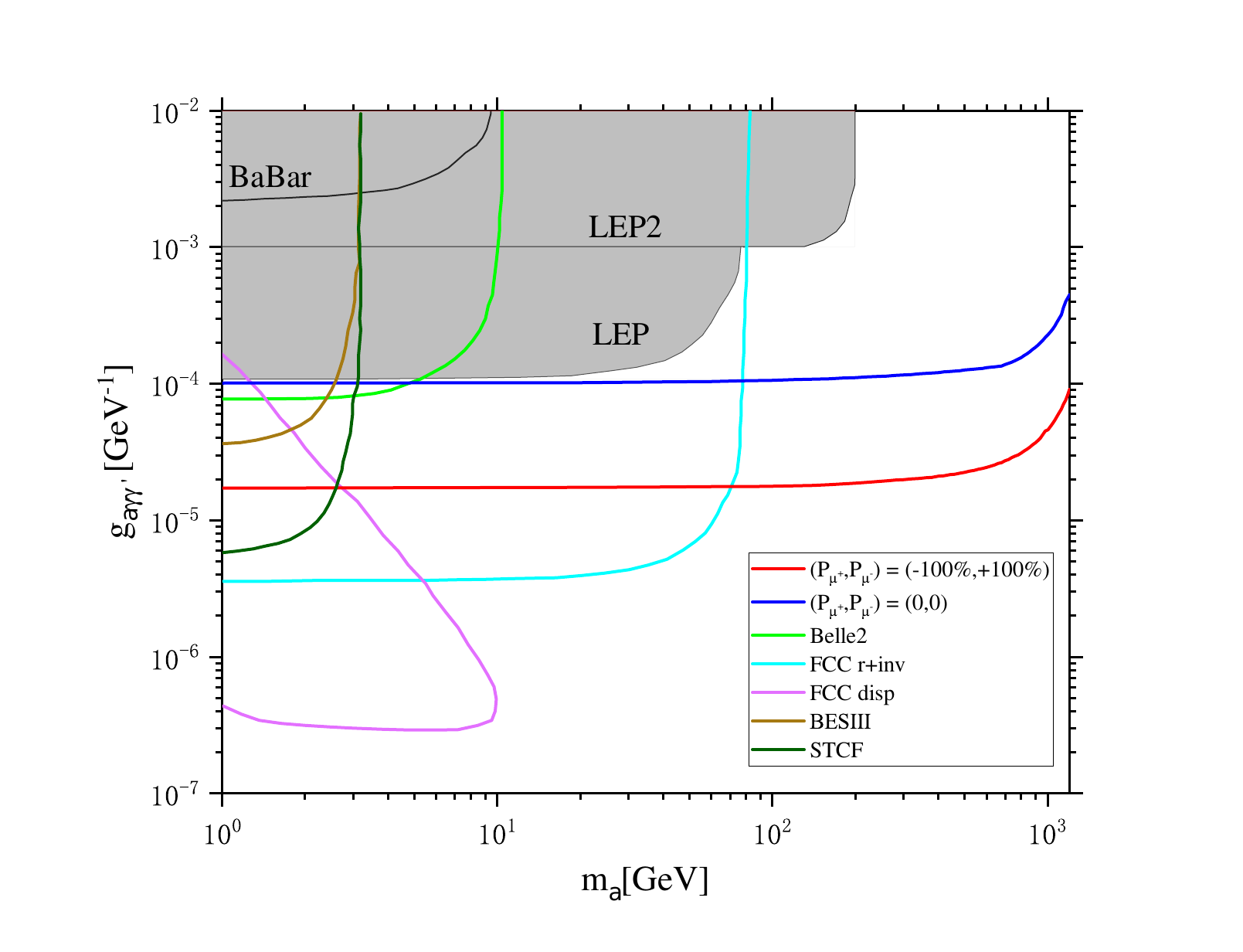}
\caption{Projected $2\sigma$ sensitivities to the $a$--$\gamma$--$\gamma'$ coupling $g_{a\gamma\gamma'}$ of the $1.5~\mathrm{TeV}$ muon collider with $\mathcal{L}=500~\mathrm{fb}^{-1}$ for the polarized $P(\mu^+,\mu^-)=(-100\%,+100\%)$ (red line) and unpolarized $P(\mu^+,\mu^-)=(0,0)$ (blue line) beam configurations, together with the current constraints from BaBar~\cite{deNiverville:2018hrc,BaBar:2008aby}, LEP~\cite{Jodlowski:2024ayf,OPAL:1994kgw,DELPHI:1996drf}, and LEP2~\cite{Jodlowski:2024ayf,DELPHI:2003dlq} and the projected sensitivities from Belle2 with $50~\mathrm{ab}^{-1}$~\cite{deNiverville:2018hrc}, FCC-ee~\cite{Jodlowski:2024ayf}, BESIII and STCF for $\epsilon_{3\gamma}^{\rm leak}=0$~\cite{Wang:2026fnr}.}
\label{fig:7}
\end{figure}

We further compare the projected $2\sigma$ sensitivities obtained in this work with the existing experimental constraints and the projected sensitivities from other collider studies, as presented in Fig.~\ref{fig:7}. The gray shaded regions represent the existing exclusion limits from BaBar~\cite{deNiverville:2018hrc,BaBar:2008aby}, LEP~\cite{Jodlowski:2024ayf,OPAL:1994kgw,DELPHI:1996drf}, and LEP2~\cite{Jodlowski:2024ayf,DELPHI:2003dlq}. The green, cyan, magenta, brown, and dark-green curves represent the projected sensitivities from Belle2~\cite{deNiverville:2018hrc}, FCC-ee with the $\gamma+\mathrm{inv}$ signature~\cite{Jodlowski:2024ayf}, FCC-ee with displaced signature~\cite{Jodlowski:2024ayf}, BESIII~\cite{Wang:2026fnr}, and STCF~\cite{Wang:2026fnr}, respectively. The red and blue curves show the projected $2\sigma$ sensitivities obtained in this work for the polarized and unpolarized beam configurations, $P(\mu^+,\mu^-)=(-100\%,+100\%)$ and $P(\mu^+,\mu^-)=(0,0)$, respectively. The projected sensitivities reach the $\mathcal{O}(10^{-5})~\mathrm{GeV}^{-1}$ level for $P(\mu^+,\mu^-)=(-100\%,+100\%)$ and the $\mathcal{O}(10^{-4})~\mathrm{GeV}^{-1}$ level for $P(\mu^+,\mu^-)=(0,0)$. Both extend beyond the regions excluded by existing experiments. Among electron--positron collider searches, Belle2 with $50~\mathrm{ab}^{-1}$ can probe the coupling at the $\mathcal{O}(10^{-5})~\mathrm{GeV}^{-1}$ level in the low-mass region. BESIII and STCF are also sensitive to relatively light ALPs, reaching the $\mathcal{O}(10^{-5})~\mathrm{GeV}^{-1}$ and $\mathcal{O}(10^{-6})~\mathrm{GeV}^{-1}$ levels, respectively, for a three-photon leakage rate of $\epsilon_{3\gamma}^{\rm leak}=0$. The projected sensitivities can be affected by a nonzero three-photon leakage rate, as discussed in detail in Ref.~\cite{Wang:2026fnr}. At higher energies, FCC-ee can reach the $\mathcal{O}(10^{-6})~\mathrm{GeV}^{-1}$ level for ALP masses in the range $1~\mathrm{GeV}\leq m_a\leq50~\mathrm{GeV}$, while for long-lived ALPs with displaced decays, the sensitivity can reach the $\mathcal{O}(10^{-7})~\mathrm{GeV}^{-1}$ level in the range $1~\mathrm{GeV}\leq m_a\leq10~\mathrm{GeV}$. Although several future colliders can achieve stronger sensitivities, the muon collider extends the exploration of the $a$--$\gamma$--$\gamma'$ interaction to higher ALP masses and therefore provides a complementary probe of the parameter space.

\section{Summary and discussion}

In recent years, extensive studies have been devoted to searching for new physics beyond the SM through both theoretical predictions and experimental searches. Among the proposed scenarios, the dark axion portal has emerged as a well-motivated extension of the ALP and dark photon frameworks. Its characteristic feature is that the effective $a$--$\gamma$--$\gamma'$ interaction provides a unique connection between the visible and dark sectors and gives rise to rich collider phenomenology. Although this interaction has been extensively studied at both electron--positron and hadron colliders, a systematic investigation at a muon collider is still lacking. Future muon colliders therefore provide a promising opportunity to probe the $a$--$\gamma$--$\gamma'$ interaction and further extend the collider exploration of the dark axion portal.

In this paper, we have investigated the collider phenomenology of the $a$--$\gamma$--$\gamma'$ interaction through the process $\mu^{+}\mu^{-}\to a\gamma'$, followed by the prompt decay $a\to\gamma\gamma'$. Assuming $m_a\gg m_{\gamma'}$ and the dark photons escaping detection, the signal is characterized by a clean final state with a single energetic and isolated photon accompanied by invisible particles. A detector-level analysis has been carried out for the $1.5~\mathrm{TeV}$ muon collider with $\mathcal{L}=500~\mathrm{fb}^{-1}$ under the beam polarizations $P(\mu^+,\mu^-)=(-100\%,+100\%)$ and the unpolarized beam configuration $P(\mu^+,\mu^-)=(0,0)$. The projected sensitivity reaches the $\mathcal{O}(10^{-5})~\mathrm{GeV}^{-1}$ level for ALP masses in the range $1~\mathrm{GeV}\leq m_a\leq1200~\mathrm{GeV}$ for $P(\mu^+,\mu^-)=(-100\%,+100\%)$, while, for $P(\mu^+,\mu^-)=(0,0)$, the projected sensitivity reaches the $\mathcal{O}(10^{-4})~\mathrm{GeV}^{-1}$ level for the same ALP mass range. Compared with the existing experimental constraints, the projected sensitivities obtained in this work indicate that both the polarized and unpolarized muon collider have the potential to probe regions of parameter space beyond the current experimental limits. These results demonstrate that a muon collider offers a complementary probe of the $a$--$\gamma$--$\gamma'$ interaction over a broad ALP mass range. Moreover, the present analysis assumes prompt ALP decays. If the ALP is sufficiently long-lived to produce displaced signatures, substantially stronger sensitivities may be achieved, making this an interesting direction for future studies.

\section*{ACKNOWLEDGMENT}

This work was partially supported by the National Natural Science Foundation of China under Grant No. 12575106 and the Cultivation Fund of Liaoning Normal University for Excellent Doctoral Dissertations~(No. YJSYB202501).


\end{document}